\documentclass[amsmath, amssymb, reprint, aps, superscriptaddress, floatfix]{revtex4-1}
\usepackage{epsfig,xcolor}
\usepackage{natbib}

\newcommand{\fsky}{f_{\rm sky}}
\newcommand{\nhat}{{\bf \hat n}}
\newcommand{\mth}{m^{\rm th}}

\newcommand{\zmax}{{z_{\rm max}}}
\newcommand{\lmax}{{\ell_{\rm max}}}
\newcommand{\fwhm}{{\rm FWHM}}
\newcommand{\nside}{\textsc{Nside}}
\newcommand{\Cl}{C_\ell}
\newcommand{\Ylm}{Y_{\ell m}}
\newcommand{\alm}{a_{\ell m}}
\newcommand{\ds}{\displaystyle}
\newcommand{\healpix}{{\tt HEALPix}\,}
\newcommand{\rdata}{r^{\rm data}}
\newcommand{\riso}{r^{\rm iso}}

\begin{document}

\title{Percent-Level Test of Isotropic Expansion Using Type Ia Supernovae}

\author{John Soltis}
\affiliation{Department of Physics, University of Michigan,  450 Church St, Ann Arbor, MI 48109-1040}

\author{Arya Farahi}
\affiliation{Department of Physics, University of Michigan,  450 Church St, Ann Arbor, MI 48109-1040}
\affiliation{Leinweber Center for Theoretical Physics, University of Michigan, 450 Church St, Ann Arbor, MI 48109-1040}
\affiliation{McWilliams Center for Cosmology, Department of Physics, Carnegie Mellon University, Pittsburgh, PA 15213-3890}

\author{Dragan Huterer}
\affiliation{Department of Physics, University of Michigan,  450 Church St, Ann Arbor, MI 48109-1040}
\affiliation{Leinweber Center for Theoretical Physics, University of Michigan, 450 Church St, Ann Arbor, MI 48109-1040}

\author{C.\ Michael Liberato II}
\affiliation{Department of Physics, University of Michigan,  450 Church St, Ann Arbor, MI 48109-1040}

\begin{abstract}
We propose and implement a novel, robust, and non-parametric test of statistical isotropy of the expansion of the universe, and apply it to around one thousand type Ia supernovae from the Pantheon sample. We calculate the angular clustering of supernova magnitude residuals and compare it to the noise expected under the isotropic assumption. We also test for systematic effects and demonstrate that their effects are negligible  or are already accounted for in our procedure. We express our constraints as an upper limit on the rms spatial variation in the Hubble parameter at late times. For the sky smoothed with a Gaussian with $\fwhm=60^\circ$, less than $1\%$ rms spatial variation in the Hubble parameter is allowed at 99.7\% confidence.
\end{abstract}

\maketitle

\textit{Introduction.}
The simplest inflationary-cosmology scenarios \cite{Guth:1980zm,Linde:1981mu,Mukhanov:1981xt,Albrecht:1982wi} generically predict that the expansion of the universe is isotropic. However, violations of statistical isotropy can  certainly be accommodated in models with additional complexity (e.g.\ \cite{Ackerman:2007nb,Pullen:2007tu,Watanabe:2009ct}) and even perturbative effects on the expansion rate in the standard cosmological model (which are, however, expected to be small, e.g.\ \cite{Bonvin:2005ps,Bonvin:2006en}). While tests of statistical isotropy of the \textit{early} universe have typically been carried out by analyzing the cosmic microwave background anisotropy maps (e.g.\ \cite{Hanson:2009gu,Bennett:2010jb,Copi:2010na,Kim:2013gka,Ade:2015hxq}), it is well worthwhile to investigate the isotropy of the \textit{late}-time universe. The latter is particularly interesting given the lack of a fundamental understanding of the physical nature of dark energy that powers the accelerated expansion of the universe. 

In this paper we present a novel test of the isotropy of cosmic expansion and apply it to current type Ia supernovae (SNIa) data. While an investigation of the isotropy of the universe using SNIa data has been carried out by numerous previous works  \cite{Antoniou:2010gw,Colin:2010ds,Campanelli:2010zx,Cai:2011xs,Mariano:2012wx,Appleby:2012as,Kalus:2012zu,Cai:2013lja,Zhao:2013yaa,Jimenez:2014jma,Appleby:2014kea,Bengaly:2015dza,Bengaly:2016amk,Lin:2015rza,Javanmardi:2015sfa,Huterer:2016uyq,Wang:2017ezt,Andrade:2017iam,Sun:2018cha,Andrade:2018eta,Deng:2018jrp}, our methodology (described below) extends these efforts. Our approach is parameter-free, robust, and explicitly independent of assumptions about the distribution of the data. We now describe the data we use, our methodology, and present the results along with estimates of the effects of systematic errors.

\medskip
\textit{Data.}
For our analysis, we use the ``Pantheon'' compilation of SNIa \cite{Scolnic:2017caz}. The Pantheon sample combines 279 SNIa $(0.03 < z < 0.68)$ from the Pan-STARRS1 Medium Deep Survey with SNIa from Sloan Digital Sky Survey (SDSS), SuperNova Legacy Survey (SNLS), and various low-z and Hubble Space Telescope samples to produce a SNIa sample of 1048 objects in the redshift range $0.01 < z < 2.26$.  The Pantheon sample was produced using the PS1 Supercal process \cite{Scolnic:2015eyc}, which determined a global calibration solution to combine 13 different SNIa samples. The latter analysis also corrects for expected biases in light-curve fit parameters and their errors using the method outlined in Ref.~\cite{Kessler:2016uwi}. 

\medskip
\textit{Methodology.} The SNIa data consists of individual magnitude
measurements $m_i\equiv m(z_i, \hat{\mathbf{n}})$, where $z_i$ is the redshift
of a supernova in the cosmic microwave background frame and corrected for
peculiar velocities \cite{Scolnic:2017caz}, and $\hat{\mathbf{n}}$ is its
location on the sky. The individual SNIa magnitude errors $\sigma_i$ are
generalized, in modern SNIa analyses, to the full covariance matrix,
$\mathbf{C}=\mathbf{S}+\mathbf{N}$, where $\mathbf{S}$ and $\mathbf{N}$ are
the signal and noise matrices, respectively. The noise matrix encodes
statistical magnitude measurement errors and covariances due to unknown fit
parameters that correlate the measurements, such as the color and stretch. The
signal matrix is nonzero at low redshift because, roughly speaking, nearby
SNIa are pulled by the same structures, resulting in correlated peculiar
velocities. At high redshift the signal matrix is nonzero mainly because of
the effects of lensing on SNIa magnitudes.

\begin{figure}[t]
\includegraphics[width=0.48\textwidth]{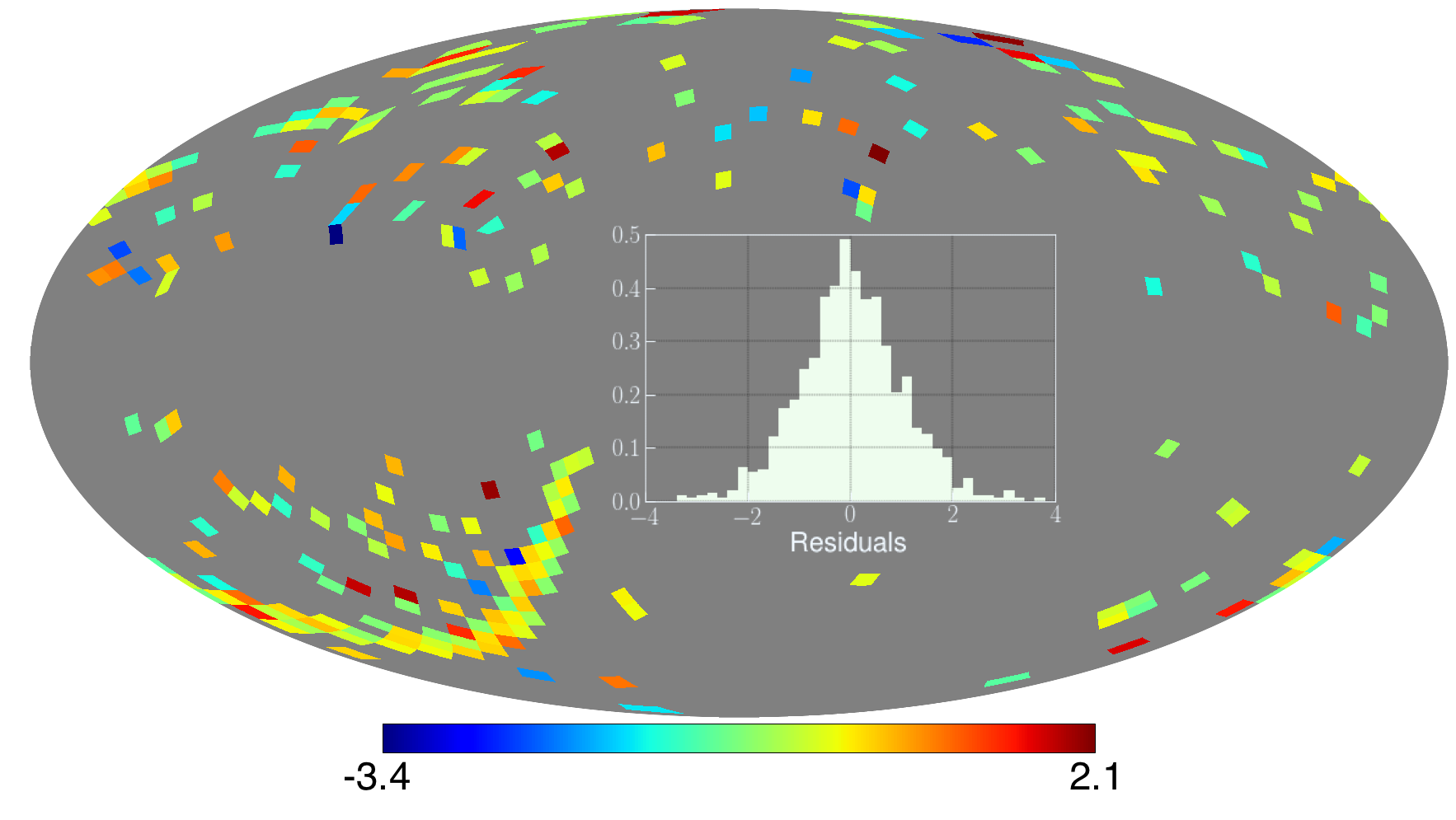}
\caption{Mollweide-projection map of SNIa magnitude residuals, defined in Eq.~(\ref{eq:res}), in Galactic coordinates and at \healpix\ resolution $\nside=16$. Each pixel contains the average of the residuals of SNIa that fall in it. The inset shows the histogram of the SNIa residuals.}
\label{fig:map}
\end{figure}

We work with  SNIa magnitude residuals divided by individual statistical errors 
\begin{equation}
\label{eq:res}
r_i\equiv \ds\frac{m_i-\mth_i}{\sigma_i}
\end{equation}
where $\mth_i=5\log_{10}[H_0d_L(z_i, \Omega_M)] + \mathcal{M}$ is the theoretically expected magnitude for an object at redshift $z_i$ and a given value of the matter density relative to critical $\Omega_M$, in the flat $\Lambda$CDM universe which we assume. Here $\mathcal{M}$ is the nuisance parameter that combines the absolute magnitude of SNIa with the Hubble constant $H_0$. The distribution of the residuals on the sky is shown in Fig.~\ref{fig:map}; the inset in the Figure shows that the residuals are approximately Gaussian-distributed.

The goal of this study is to put an upper limit on the value of the ``signal'' in  the distribution of the SNIa magnitude residuals caused by violations of statistical isotropy. Since we are searching for the excess signal with respect to statistical noise,  we  choose to normalize the magnitude residuals in Eq.~(\ref{eq:res}) by the diagonal statistical  
error $\sigma_i$, and not elements of the full (signal plus noise) covariance $\mathbf{C}$. Note that the statistical measurement error constitutes the majority of the contribution to the diagonal of the noise covariance, $\sigma_i^2\simeq 0.99 N_{ii}$, thus we are effectively dividing by the square roots of the latter. 
The total signal in the clustering of residuals also has guaranteed contributions from the peculiar velocities of the SNIa (which are correlated because the velocity's origin is the gravitational pull of the nearby large-scale structures), and from the systematic uncertainty in the precise values of the cosmological parameters $\Omega_M$ and $\mathcal{M}$ which are required to calculate $m^{\rm th}(z)$. We demonstrate below that these additional contributions to the signal are very small compared to the noise level in the SNIa data and can be ignored.

\begin{figure}[t]
\includegraphics[width=0.48\textwidth]{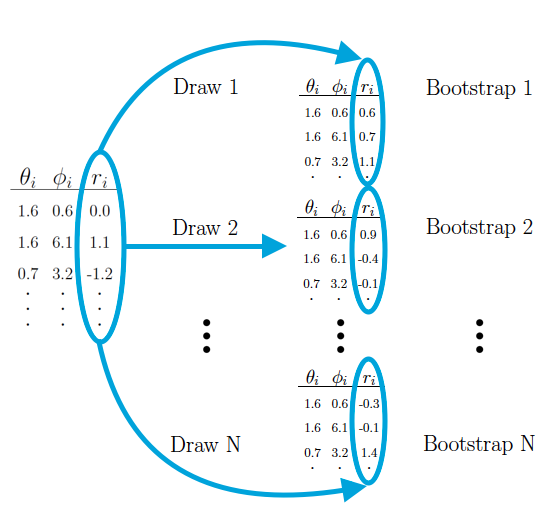}
\caption{Illustration of the process to estimate the noise and its uncertainty in our measurements. In each bootstrap, the magnitude residual of each SNIa is replaced with one drawn from the full set of SNIa residuals in the Pantheon sample. We marginalize over the cosmological parameters by performing this bootstrap analysis for values of $\Omega_M$ and $\mathcal{M}$ drawn from the joint posterior distribution of the cosmological analysis of the Pantheon data. }
\label{fig:bootstrap}
\end{figure}

To estimate the signal power spectrum, we pixelate the sky using the \healpix \cite{Gorski:2004} resolution parameter $\nside$. Our fiducial analysis is done at $\nside=16$, though we also compare with results at $\nside=$ 8, 32, and 64, finding good agreement. Each pixel has a side of roughly $60^\circ/\nside$. Because more than one SNIa may lie in a given pixel, we choose to take the average of the residuals in the pixel. Thus, the value of the $j^{\rm th}$ pixel is given by $p_j = (\sum_{r_i\in p_j} r_i)/n_j$, where $n_j$ is the number of SNIa in that pixel and the sum goes over the residuals located in the pixel. 
We now outline how the data power spectrum is computed and then discuss our noise estimation approach.

 To calculate the angular power spectrum of the map of residuals, $\Cl$, we employ a pseudo-$\Cl$ estimator which, given the small sky coverage of the SNIa (at $\nside=16$, the fractional sky coverage is only $\fsky\simeq 0.07$), is much more practical than the maximum-likelihood estimators which try to recover the full-sky signal. To get the pseudo-$\Cl$ we adopt the function {\tt anafast} in \healpix. Given the significant variation of pixel occupancy by SNIa, it is crucial to weight each pixel by the number of objects in it; this guarantees that large-angle (low-$\ell$) $\Cl$ will not depend on the pixelation as long as the latter is finer than the scales we wish to probe. Our angular power spectrum is given by the usual pseudo-$\Cl$ formula
\begin{equation}
      C_\ell = \frac{1}{2\ell+1}\sum_{m=-\ell}^\ell |\alm |^2,
\end{equation}
with the harmonic expansion of the residuals that applies weight to the pixels
\begin{equation}
\alm \equiv 
\frac{\int r(\nhat)W(\nhat) \Ylm^*(\nhat)\,d^2\nhat}
{\int W(\nhat)\, d^2\nhat /(4\pi)}
\label{eq:alm}
\end{equation}
where the integral is typically discretized as the sum over the pixels whose centers are in directions $\nhat$ and which have areas $d^2\nhat$. Here $r(\nhat)$ is the mean residual in a pixel in the direction $\nhat$, while the weight $W(\nhat)$ is given by the number of objects in that pixel. The denominator in Eq.~(\ref{eq:alm}) evaluates to  $\fsky\langle W_{\rm pix}\rangle$, where $\langle W_{\rm pix}\rangle$ is the average number of SNIa per pixel. Note that the overall normalization of the $C_\ell$ is not important for comparing the angular spectrum of our sky to that of the bootstrapped sample since the two have the same normalization, but it \textit{is} important when we quote limits on the statistical isotropy of the expansion rate. 

Having obtained the angular power spectrum of the SNIa, we then produce the distribution of angular power spectra that would be expected in an isotropic universe, which in turn is given by the clustering noise and no signal. Because the number of SNIa varies significantly from pixel to pixel, the computation of noise in the angular clustering of the SNIa is analytically intractable, and would be so even under the simplified assumption of Gaussian noise. To get a reliable upper bound on the background anisotropy it is crucial to estimate the clustering noise, as well as its uncertainty, directly from the data. 
To address this, we employ a simple, non-parametric bootstrap approach whose principal advantage is that the noise level can be directly estimated from data. Additionally, this approach does not make assumptions about the statistical distribution of residuals, e.g.\ whether it is Gaussian. At the location of each SNIa we draw, with replacement, a residual from the distribution of the residuals of all SNIa. Having done this for all SNIa, we have a randomized  realization of the residuals --- hence one isotropic-universe bootstrap --- which we refer to as $\riso$. To estimate the statistical uncertainty due to the finite number of SNIa, we repeat this procedure, which is illustrated in Fig.~\ref{fig:bootstrap}, 1,000,000 times. 
At every 1,000-th bootstrap, we also draw the cosmological parameters $\Omega_M$ and $\mathcal{M}$ from their posterior distribution obtained using our cosmological analysis of the Pantheon SNIa, and then reevaluate the residuals. We do this to account for the imperfect knowledge of these parameters. Finally, we calculate the angular power spectrum for each bootstrapped realization. This procedure produces a range of values of $C_\ell$ expected in an isotropic universe for a discrete realization of SNIa. 

\begin{figure}[t]
\includegraphics[width=0.48\textwidth]{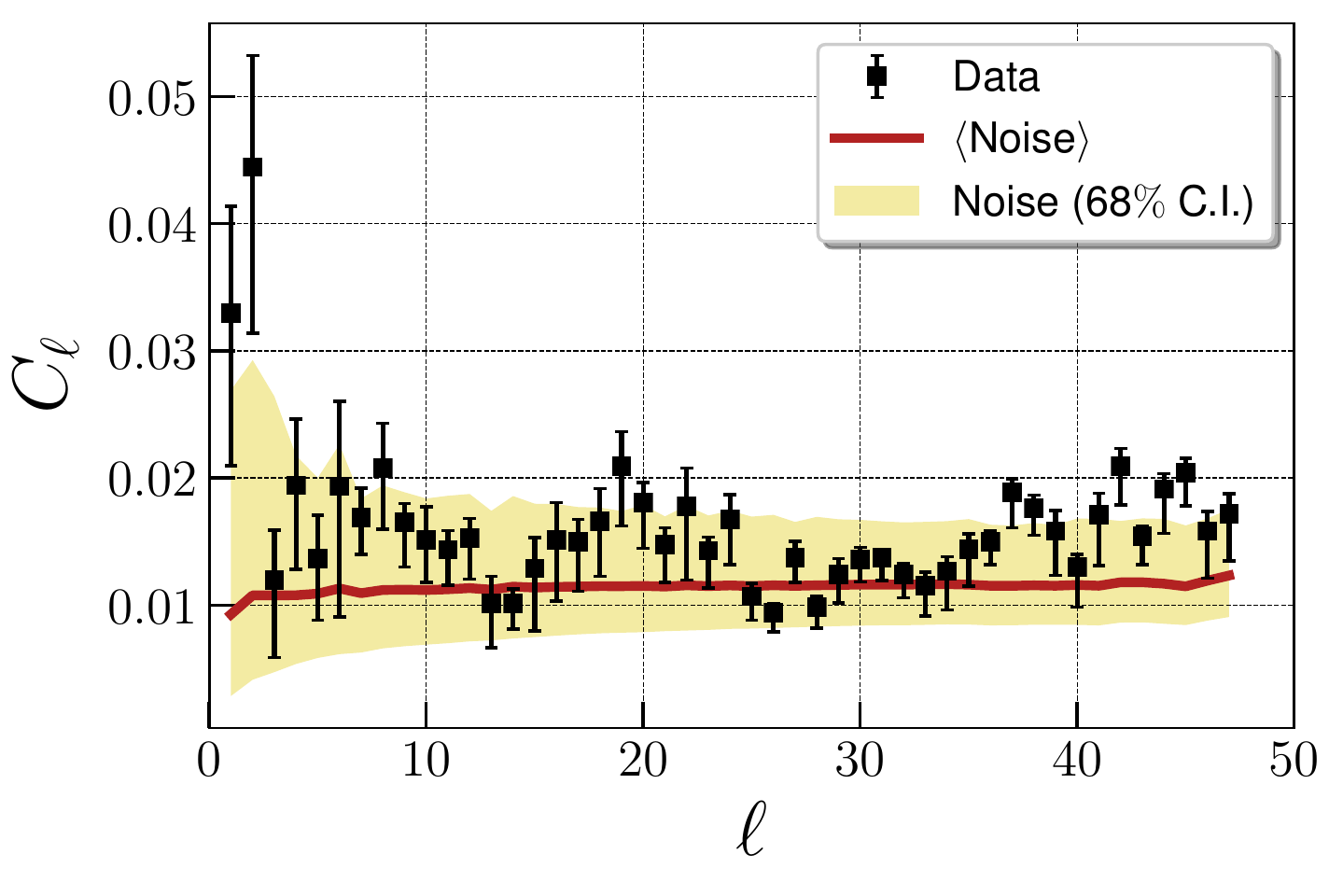}\vspace{-0.3cm}
\caption{Angular power spectrum of SNIa magnitude residuals in the Pantheon sample (black data points; errors show the effect of uncertain cosmological parameters). The near-horizontal thin red line shows the mean values of $C_\ell$ expected due to statistical fluctuations (noise) in an isotropic-universe, while the yellow region shows the 68\% confidence interval uncertainty in it.  See text for details.}
\label{fig:Cl}
\end{figure}

\medskip
\textit{Results.}
Figure \ref{fig:Cl} shows the angular power spectrum of the pixelated SNIa average residuals; the black error bars indicate the effect of the uncertain knowledge of cosmological parameters, corresponding to the uncertainty in theoretical magnitudes $\mth_i$. We also show the distribution of $C_\ell$ of bootstrap rearrangements of residuals on the sky: the dark thin red curve shows the mean value of the $\Cl$ due to noise and calculated from our million bootstraps, while the yellow region around it shows the 68\% uncertainty in this distribution. 
Here and in what follows we show results for $\nside=16$. We have checked that our results are basically unchanged for $\nside=8$, 32, and 64.

Figure \ref{fig:Cl} indicates that the angular power spectrum of the Pantheon data appears consistent with the isotropic assumption given by the bootstraps, which encode the theoretical expectation of clustering noise and no signal. A simple chi-squared test confirms this; we calculate the quantity
\begin{equation}
  \chi^2 = 
  \left (C_\ell-\bar{C}_\ell^{\rm boot}\right )
  (M^{-1})_{\ell\ell'}
  \left (C_{\ell'}-\bar{C}_{\ell'}^{\rm boot}\right ),
\label{eq:chisq}
\end{equation}
where the sum over the multipoles is implied and where $C_\ell$ and $\bar{C}_\ell^{\rm boot}$ correspond to the data and the mean of the bootstraps respectively. The coupling matrix $M_{\ell\ell'}= \langle(C_{\ell}^{\rm boot} - \bar{C}_{\ell}^{\rm boot}) (C_{\ell'}^{\rm boot} - \bar{C}_{\ell'}^{\rm boot}) \rangle$ is calculated directly from the bootstraps, and is non-diagonal because the SNIa do not cover the full sky.  We find that $\chi^2/{\rm dof} = 1.41$ for a total of 48 degrees of freedom, being a little under the 2-$\sigma$ level (the significance is even smaller for other pixelations we looked at, $\nside=8, 32$, and 64).  Therefore, the null hypothesis of an isotropic expansion rate cannot be rejected, and this preliminary test of the isotropy is passed.  We now turn to a more quantitative interpretation of our results.

We would like to get additional insight on how well our data constrain the
isotropy in an as model-independent way as possible. To that effect, we
consider a (redshift-independent) fractional variation in the expansion rate
at $z\lesssim 1$
\vspace{-0.3cm}
\begin{equation}
  \delta_H(\nhat)\equiv \frac{\delta H}{H}(\nhat)\ll 1.
  \label{eq:deltah}
\end{equation}
Propagating it through to the magnitude and labeling the perturbed magnitudes
with a tilde, it follows that $\widetilde{m}_i(\nhat) = m_i(\nhat) + (5/\ln
10)[1-\delta_H(\nhat)]$.  The variance of the residuals subject to such isotropy breaking, $\langle (\rdata)^2\rangle$, is then 
\begin{equation}
  \ds\langle (\rdata)^2 \rangle = 
  \left (\frac{5}{\ln 10}\right )^2
  \left\langle \left (\frac{\delta_H}{\sigma}\right )^2\right\rangle + N(\riso) 
  \label{eq:expect_rsq}
\end{equation}
where the noise term is given by the variance expected in an isotropic universe due to chance statistical fluctuations (as well as any systematic uncertainties), $N(\riso)\equiv \langle (\riso)^2\rangle$.
Here we have assumed no correlation between the random fluctuations in the \textit{isotropic} residuals $\riso(\nhat)$ and the isotropy breaking $\delta(\nhat)$, which is justified given their completely different origins.

Conveniently, our angular power spectrum measurement can be converted to the variance of the residuals on the sky via
\vspace{-0.2cm}
\begin{equation}
  \mathrm{Var}(\rdata) \equiv \langle (\rdata)^2\rangle
  = \sum_{\ell=1}^\lmax \frac{2\ell+1}{4\pi} C_\ell
  \label{eq:var_vs_Cl}
\end{equation}
where we define the sum in the range that we measured the multipoles, $\ell\in[1, \lmax]$. 

\begin{figure}[t]
\includegraphics[width=0.48\textwidth]{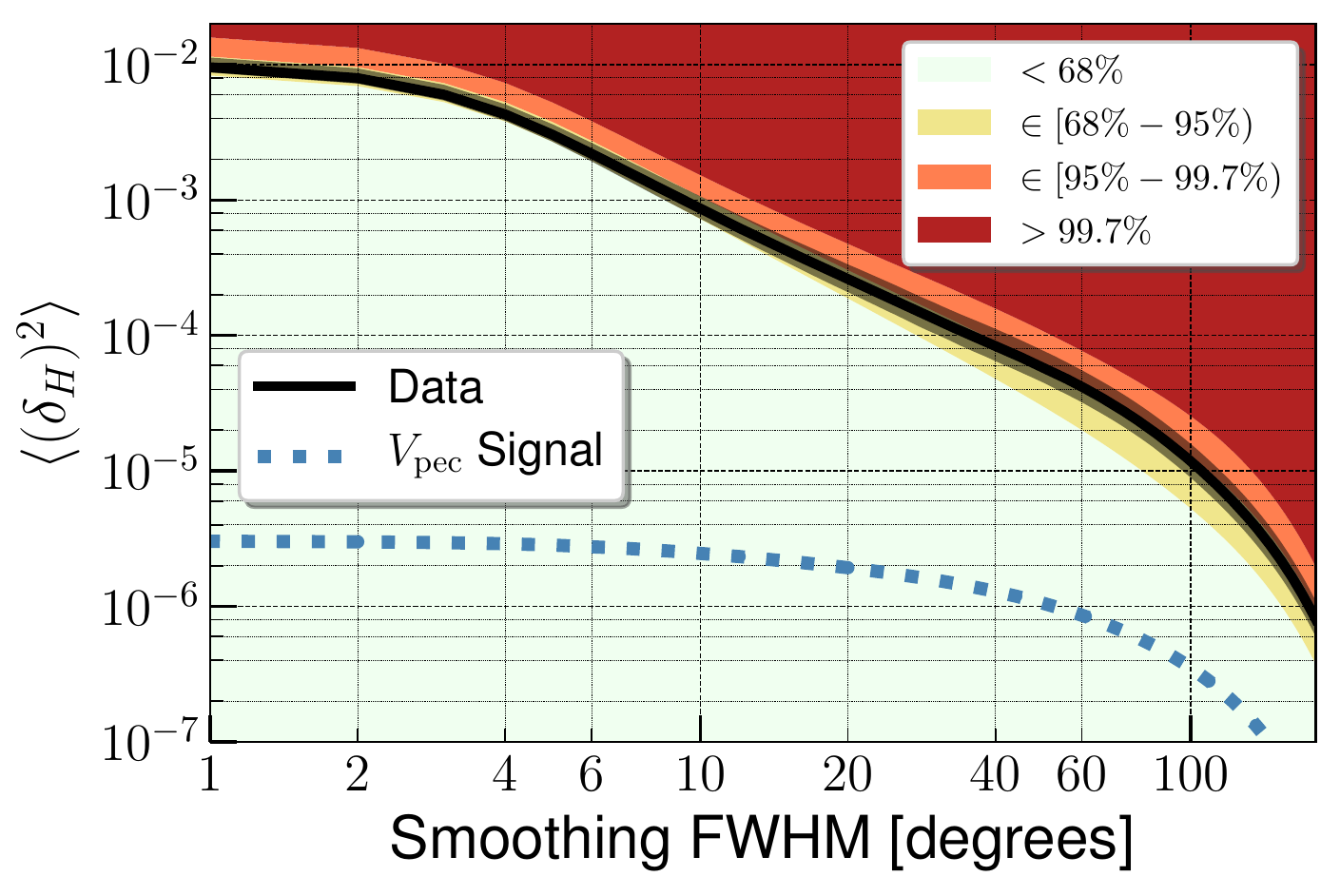}
\caption{Variance in the expansion rate derived from the clustering of SNIa
  residuals on the sky. The thick solid curve shows the signal in the
  data. The color-coded regions present the 68\%, 95\%, and 99.7\% upper
  limits derived from the bootstraps, which represent expectations for the
  isotropic distribution of SNIa residuals. The thin grey band around the
  black curve shows the effect on the variance of the data due to the
  uncertainty in the cosmological-parameter values. Note that the curves
  flatten out around 3 degrees as expected, since this is the resolution limit
  imposed by the pixelation with $\nside=16$. The dotted curve shows the
  expected contribution to the signal of the SNIa's peculiar velocities.  }
\label{fig:var_limits}
\end{figure}

Because both the data and bootstrap variance increase with maximum multipole, each going roughly as $(\lmax)^2$ for a flat power spectrum, the results depend on the resolution of the map. Physically this makes sense, as a finer-resolution map allows for additional, smaller-scale anisotropic spatial modes that can lurk just below the detection level, and hence leads to a weaker overall constraint on the breaking of isotropy. To address this in a way that is both quantitative and physically motivated, we smooth the residual maps. It is sufficient to do this in multipole space; we apply a Gaussian beam $b_\ell$ that depends on the desired smoothing FWHM; the effect on the angular power spectrum is then $\Cl^{\rm smoothed}=b_\ell^2\Cl$. 

We are finally ready to recast our results as limits on the variation of the Hubble parameter at the present time.  From Eq.~(\ref{eq:expect_rsq}), we first cast the variance in the observed residuals as the rms Hubble parameter variation. If the noise term could be neglected, the rms Hubble parameter variation would be, according to Eq.~(\ref{eq:expect_rsq})
\begin{equation}
  (\delta_H)_{\rm rms} \simeq 
  \left (\frac{\ln 10}{5}\right ) 
  \langle \sigma^{-2}\rangle^{-1/2}\,
  \sqrt{\mathrm{Var}(\rdata)}
  \label{eq:dH_from_var}
\end{equation}
 where $\langle \sigma^{-2} \rangle^{-1/2}\simeq 0.13$ is the inverse-square-weighted intrinsic dispersion of Pantheon SNIa. Given (see below) that we do not observe the evidence for a ``signal''  --- a larger variance of the residuals than that expected in the statistically isotropic universe --- the variance evaluated on isotropic bootstraps alone, $N(\riso)$, will serve to produce an \textit{upper limit} on $(\delta_H)_{\rm rms}$ as per Eq.~(\ref{eq:dH_from_var}).

The principal results are shown in Fig.~\ref{fig:var_limits}. Here we show the 68\%, 95\%, and 99.7\% upper limits on the expansion rate variance $\langle\delta_H^2\rangle$ from the bootstraps, along with the variance computed from the Pantheon SNIa data, both as a function of the smoothing scale. The grey band shows the effect on the variance of the data due to the uncertainty in the cosmological-parameter values. As mentioned above, coarser smoothing implies more stringent constraints and vice versa. The dotted line shows the guaranteed signal given by the peculiar velocities of the SNIa in the Pantheon sample, calculated using the formalism in \cite{Huterer:2016uyq}; the effect is very small and can be neglected. Note also that our limits include the marginalization over the uncertainty in the cosmological parameters since, in generating the bootstraps, we draw $\Omega_M$ and $\mathcal{M}$ (and hence values of theoretical magnitudes $\mth_i$ and the corresponding residuals $r_i$) from the posterior distribution of these two parameters obtained from the cosmological analysis of the Pantheon SNIa dataset.

Figure \ref{fig:var_limits} indicates no evidence for breaking of the assumption of isotropy and leads to quantitative limits on its breaking. For example, for the FWHM = 60 degree smoothing, the 99.7 percentile upper limit calculated from the bootstraps is $\mathrm{Var}(\riso)=0.0215$, and thus
\begin{equation}
\left [(\delta_H)_{\rm rms}\right ]^{\rm FWHM =60^\circ} < 0.009 \quad \mbox{(99.7\% CL)},
\end{equation}
 or a $\lesssim$ 1\% constraint on isotropy of the expansion at large angular scales. To stress-test the dependence of our constraints on data selection we have repeated the analysis with only SNIa out to maximum redshift $\zmax\in\{0.1, 0.2, 0.5, 1.0\}$. While both the variance in the data and the noise limits increase with decreasing $\zmax$, reproduction of Fig.~\ref{fig:var_limits} in these cases reveals results qualitatively similar to our fiducial analysis, with no evidence for breaking of the assumption of statistical isotropy. 

\medskip      
\textit{Conclusions.} We have proposed and carried out a non-parametric test of the statistical isotropy of the late-time universe. Our test utilizes the Pantheon set of just over a thousand type Ia supernovae, whose clustering we measure by evaluating the angular power spectrum of the SNIa residuals relative to the best-fit cosmological model. We use a novel --- to these tests of isotropy --- and simple method of estimating the noise that describes the clustering expected in the isotropic case by bootstrapping the spatial distribution of the SNIa residuals.

To further quantify and summarize our findings, we evaluate the variance of the residuals i.e.\ calculate their zero-lag correlation function, and express the results in terms of constraints on the rms spatial variation of the expansion rate $(\delta_H)_{\rm rms}\equiv \langle (\delta H(\nhat)/H)^2\rangle^{1/2}$, where the latter is constrained at $z\sim 0.3$ where SNIa have the most constraining power. Because this quantity increases as smaller spatial scales are probed we explicitly smooth the angular power spectrum, evaluating the rms variation as a function of the smoothing scale. 

Our results show no evidence for breaking of statistical isotropy in the Pantheon sample and, for the first time to our knowledge, constrain it at better than the 1\% level at large spatial scales (smoothing FWHM $\gtrsim 60^\circ$); see Fig.~\ref{fig:var_limits}.

We pay particular attention to the control and understanding of systematic errors. Our analysis choices ensure that our results do not depend on the pixelation of the map of SNIa residuals. We explicitly account for the uncertainty in the values of the cosmological parameters used to calculate the residuals and for the fact that there is a guaranteed signal of anisotropy due to the peculiar velocities of nearby objects; both effects are small and we explicitly marginalize over the former. We also find no qualitative change in our results when we restrict the range of redshifts of the SNIa in the Pantheon sample.

Our analysis does not assume the Gaussianity of the SNIa residuals, although the latter does approximately hold. We do assume the $\Lambda$CDM cosmological model; this is justifiable given the lack of evidence for its extensions (e.g.\ \cite{Aghanim:2018eyx}).

Our test therefore constrains the isotropy of the expansion rate at $z\lesssim 1$ at the $\sim 1\%$ level at the largest angular scales and complements the corresponding (though 2-3 orders-of-magnitude stronger) tests in the early universe.

\medskip
\textit{Acknowledgments.}
JS and DH are supported by NASA under contract 14-ATP14-0005; DH is also supported by DOE under Contract No. DE-FG02-95ER40899. AF is supported by a McWilliams Postdoctoral Fellowship.

\bibliography{myref}

\end{document}